\documentclass[aps,prX,twocolumn,superscriptaddress]{revtex4}

\usepackage{fancyhdr}     
\usepackage{amsfonts}
\usepackage{amsmath}
\usepackage[latin1]{inputenc}
\usepackage{graphicx,graphics,amssymb,times}
\usepackage{color}
\usepackage{natbib}
\usepackage{comment}

\newcommand{\vecx}[1]{\boldsymbol{\mathrm{#1}}}

\begin{document}


\title{Testing for entanglement with periodic coarse-graining}

\begin{abstract}
Continuous variables systems find valuable applications in quantum information processing. To deal with an infinite-dimensional Hilbert space, one in general has to handle large numbers of discretized measurements in tasks such as entanglement detection. 
Here we employ the continuous transverse spatial variables of photon pairs to experimentally demonstrate novel entanglement criteria based on a periodic structure of coarse-grained measurements. The periodization of the measurements allows for an efficient evaluation of entanglement using spatial masks acting as mode analyzers over the entire transverse field distribution of the photons and without the need to reconstruct the probability densities of the conjugate continuous variables. Our experimental results demonstrate the utility of the derived criteria with a success rate in entanglement detection of $\sim60\%$ relative to $7344$ studied cases. 
\end{abstract}

\author{D. S. Tasca}
\email[electronic address: ]{dan.tasca@gmail.com}
\address{Instituto de F\'isica, Universidade Federal Fluminense, Niter\'oi, RJ
24210-346, Brazil}
\address{Instituto de F\'{\i}sica, Universidade Federal do Rio de Janeiro, Caixa Postal 68528, Rio de Janeiro, RJ 21941-972, Brazil}

\author{{\L}ukasz Rudnicki}
\address{Max-Planck-Institut f{\"u}r die Physik des Lichts, Staudtstra{\ss}e 2, 91058
Erlangen, Germany}
\address{Center for Theoretical Physics, Polish Academy of Sciences, Aleja Lotnik\'ow 32/46, 02-668 Warsaw, Poland}

\author{R. S. Aspden}
\address{SUPA, School of Physics and Astronomy, University of Glasgow, Glasgow, G12 8QQ, United Kingdom}

\author{M. J. Padgett}
\address{SUPA, School of Physics and Astronomy, University of Glasgow, Glasgow, G12 8QQ, United Kingdom}

\author{P. H. Souto Ribeiro}
\address{Departamento de F\'{\i}sica, Universidade Federal de Santa Catarina, Florian\'{o}polis, SC 88040-900, Brazil}

\author{S. P. Walborn}
\address{Instituto de F\'{\i}sica, Universidade Federal do Rio de Janeiro, Caixa Postal 68528, Rio de Janeiro, RJ 21941-972, Brazil}

\maketitle

\date{\today}

\section{Introduction}
The efficient preparation and manipulation of high-dimensional quantum systems allow for the processing of large amounts of quantum information. In addition to advantages, such as increased transmission rates, a number of interesting fundamental aspects related to quantum entanglement \cite{horodecki09,guhne09}, non-locality \cite{Brunner14} and contextuality \cite{Amaral18} only become evident with the use of quantum systems of high-dimension.
\par
The transverse spatial modes of single or entangled photons constitute an interesting experimental platform for the study of high-dimensional quantum systems \cite{Walborn10}. In principle, spatial degrees of freedom (DOF) are described by an infinite set of modes: the transverse position or momentum modes provide the continuous variable (CV) description of the spatial DOF, while infinite-dimensional discrete basis can also be explored by using, for instance, the orbital angular momentum \cite{Allen92,Padgett17}, and radial modes \cite{Karimi14a,Zhou17}.  In the latter approach, a finite dimensional space can be achieved by simply isolating a subset of the entire set of spatial modes. For these $d$-dimensional states, entanglement detection \cite{Walborn04a,Leach09,Dada11a,Salakhutdinov12,Giovannini13,Hiesmayr13,Krenn14} is performed with tools developed for finite dimensional quantum systems \cite{Spengler12,Erker17}.   
\par
In the CV regime, on the other hand, the observation of spatial entanglement typically requires the measurement of the joint distributions for the position and momenta of the photons \cite{edgar12,Moreau12,Moreau14,Defienne18}, from which criteria devoted to CV systems are used to test for non-separability \cite{Duan00,Mancini02,Shchukin05}, EPR \cite{Reid89,Reid09,Walborn11} correlations and ``steering" \cite{Wiseman07}. Albeit fundamentally continuous, real-world experiments are subject to the coarse-graining imposed by the detector resolution \cite{Rudnicki12a}, as well as the limited detector range \cite{Ray13a}.  Both of these issues can lead to false positives concerning detection of quantum correlations \cite{Tasca13a,Ray13a}.  Even though typical CV entanglement and EPR criteria have been adapted for coarse-grained measurements \cite{Tasca13a,Schneeloch13a}, the experimental assessment of the position and momentum correlations typically require a large number of measurements.
\par
A number of techniques have been applied in effort to reduce the number of measurements necessary to identify entanglement in these systems, while maintaining the full range of detection events.  
For instance, Howland {\it et al.} \cite{Howland13} have shown that the use of compressive sensing techniques allows for the reconstruction of the SPDC joint detection probabilities with an efficiency improvement over a raster scanning procedure with equivalent resolution. 
In this case, nevertheless, the identification of entanglement is still bound to the application of the typical CV entanglement criteria \cite{Adesso07}, based--for example--on the evaluation of the moments \cite{Duan00,Mancini02,Shchukin05} or entropy \cite{Walborn09,Saboia11} of the reconstructed distributions.  
\par
In the present contribution, we develop and test experimentally convenient entanglement criteria based on periodic coarse-grained measurements. 
We start by establishing an uncertainty relation (UR) for the localization of a single quantum particle in a periodic array of position and momentum bins. The developed UR is expressed in terms of the {\it cross-correlation} function between a periodic analyzer and the distributions for the position and momentum variables. We then use this UR to build entanglement criteria devoted to bipartite CV quantum systems and apply them to test for spatial entanglement of photon pairs from spontaneous parametric down-conversion (SPDC). In this optical setup, the developed criteria are experimentally accessible via the joint transmission of the photons through periodic apertures playing the role of spatial-mode analyzers. This measurement strategy using {\it periodic spatial mask analyzers} acting over the entire transverse field structure of the photons enables approximately uniform single-photon and higher coincidence detection rates than the traditional binning with single apertures, thus yielding better signal-to-noise ratio.
In addition, the periodicity of the spatial masks used in position ($T_x$) and momentum ($T_p$) measurements work as free parameters that can be independently tuned to optimize the entanglement detection. In our experiment, we tested $7344$ different combinations of spatial mask geometries, achieving a success rate in entanglement detection of about $60\%$.
\par
This paper is structured as follows: in sections \ref{sec:theory} and \ref{sec:UR}, we provide the theoretical background necessary for the development of our entanglement criteria, which is derived in section \ref{sec:EntCriteria}.  Our experimental scheme and measurements with the periodic spatial masks are described in section \ref{sec:experiment}, and  the analysis of our experimental data with the derived entanglement criteria is presented in section \ref{sec:results}.  We provide concluding remarks in section \ref{sec:conclusion}.
\section{Preliminaries}
\label{sec:theory}

\subsection{Optical Fourier Transform}
\begin{figure}[t]
\begin{center}
\includegraphics[width=86mm]{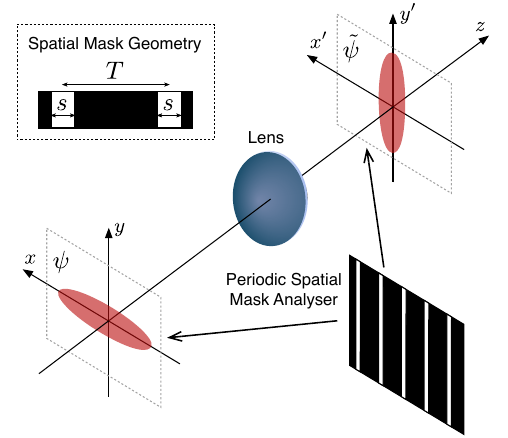}
\caption{Paraxial propagation of a single-photon through an optical Fourier transform system. The transverse field distribution of the photon at the front ($\psi$) and back ($\tilde{\psi}$) focal plane of the lens is illustrated in red. The transmission probability of the photon through a spatial mask analyzer based on a periodic aperture is used to probe the photon's transverse structure in position or momentum domain. The inset illustrates the geometry of the spatial mask analyzer.}
\label{Fig:OpticalFourier}
\end{center}
\end{figure}

The paraxial propagation of a monochromatic single-photon through a lens system implementing an optical Fourier transform is illustrated in Fig. \ref{Fig:OpticalFourier}. The propagation direction is along the positive $z$-axis and we denote $\psi(\vecx{x})=\langle \vecx{x} | \psi\rangle$ the input transverse field distribution at the front focal plane of the lens. We thus consider an input pure single-photon whose quantum state in position representation is
\begin{equation}\label{PhotonState}
|\psi \rangle=\int d\vecx{x} \, \psi(\vecx{x}) |\vecx{x}\rangle,
\end{equation}
where  $\vecx{x}=(x,y)$ is the transverse coordinate at the input plane. We also introduce the function $\tilde{\psi}$ as the field distribution at the back focal plane of the lens (the Fourier plane). The transformation connecting these two wavefunctions is
\begin{equation}\label{OpticalFourier}
 \tilde{\psi}(\vecx{x}') =\frac{1}{2\pi \alpha} \int d^2 \vecx{x} \, \psi(\vecx{x}) e^{-i \vecx{x} \cdot (\vecx{x}'/\alpha) },
\end{equation}
where $\vecx{x}'=(x',y')$ represents the transverse spatial coordinate at the Fourier plane and $\alpha=f\lambda/2\pi$ is a constant related to the optical system: $\lambda$ is the wavelength of the photon and $f$ the focal length of the lens. It is straightforward to recognize that the Fourier transformed field distribution $\tilde{\psi}$ maps the transverse structure of the input photon in momentum representation, $\phi(\vecx{p})=\langle \vecx{p} | \psi\rangle$:    
\begin{equation}\label{MomRep}
\tilde{\psi}(\vecx{x}') \propto\phi( \vecx{p}).
\end{equation}
Eq. \eqref{MomRep} implicitly assumes (we set $\hbar=1$) the relation $\vecx{x}'=\alpha \vecx{p}$ between the transverse momentum component $\vecx{p}=(p_x,p_y)$ of the photon at the input plane and the spatial coordinate at the Fourier plane. 

\subsection{Periodic Spatial Mask Analyzers}
We consider spatial mask analyzers given by periodic aperture functions $M$ whose transmittance obey $|M|^2=M$, so that $M$ assumes either the value of $0$ or $1$ across the whole transverse plane. Over a single transverse dimension, such apertures are described by periodic square waves with two independent parameters: the periodicity `$T$' and an extra spatial parameter `$s$' that we name the {\it bin width}. Explicitly, we define the periodic spatial mask analyzer as
\begin{equation} 
\label{Mdef} 
M(x;T,s)=\left\{ \begin{array}{ccc}   1, & \, 0 \leq x  {\rm \,(mod \,}  T {\rm)}  <  s \\   0, &  {\rm otherwise}  \end{array} \right. .
\end{equation} 
For a given choice of $T$ and $s$, the periodic aperture function is uniquely specified by \eqref{Mdef}, provided that the mask's origin is fixed. In general, we allow extra displacement parameters representing
positioning degrees of freedom setting the origin. It will be useful to consider the Fourier series expansion of \eqref{Mdef}:
\begin{equation}\label{MFourier}
M(x;T,s)= \sum_{n \in \mathbb{Z}} c_n(\tau s) \, e^{i  n \tau  x},
\end{equation}
where $\tau = 2\pi/T$, and 
\begin{equation}\label{cn}
c_n(\kappa)= \frac{i}{2\pi n}  \left( e^{-i n \kappa} -1\right),
\end{equation}
are the coefficients of the Fourier expansion. 

In Fig. \ref{Fig:OpticalFourier} we illustrate the periodic spatial mask analyzer over the single transverse degree-of-freedom $x$ ($x'$). We thus work with a single pair of conjugate variables satisfying the commutation relation $[\hat{x},\hat{p}]=[\hat{x},\hat{x}']/\alpha=i$. The periodic analyzer is inserted in the path of the photon, either in the front or back focal plane of the lens, and we allow distinct parameters `$T$' and `$s$' for the apertures used to probe the position ($T_x$, $s_x$) and momentum ($T_p$, $s_p$) distributions. We name it a {\it symmetric} arrangement whenever $T_x/s_x=T_p/s_p\equiv d$, but in general any combination of periodic analyzers is allowed -- note that since the transverse momentum variable $p$ is mapped to the Fourier coordinate $x'$ by means of the scaling factor $\alpha=f\lambda/2\pi$, the individual choice of the periodicity and bin width is irrelevant in a symmetric arrangement as long as the ratio $T_x/s_x=T_p/s_p$ is preserved. We further define $T_{x'}=\alpha T_p$ and $s_{x'}=\alpha s_p$ as the physical spatial parameters (units of length) of the analyzer used in the Fourier plane of the lens, whereas $T_p$ and $s_p$ express the associated quantities in momentum domain (units of inverse length). From now on, unless specified, we adopt the pair $x$ and $p$ in our analysis.

Our figure of merit for the development of the UR and entanglement criteria is the photon transmission probability  through the periodic analyzer. Denoting $P(\vecx{x})=|\psi(\vecx{x})|^2$ and  $\tilde{P}(\vecx{p})=|\phi(\vecx{p})|^2$, these transmission probabilities are
\begin{subequations}\label{TProb}
\begin{equation}\label{TProbX}
\mathcal{P}(\xi_x)=\int_{\mathbb{R}} d x \, M (x - \xi_x;T_x,s_x) P(x),
\end{equation}
\begin{equation}\label{TProbP}
\tilde{\mathcal{P}}(\xi_p)=\int_{\mathbb{R}} d p \, M (p - \xi_p;T_{p},s_{p}) \tilde{P}(p),
\end{equation}
\end{subequations}
where $P(x)=\int_{\mathbb{R}} dy P(\vecx{x})$ and $\tilde{P}(p)=\int_{\mathbb{R}} dp_y \tilde{P}(\vecx{p})$ represent the marginal  probability distributions along the relevant degree of freedom. In Eqs. \eqref{TProb}, the parameters $\xi_x$ and $\xi_p$ describe the transverse displacement of the spatial mask analyzer. The transmission probabilities \eqref{TProb} can be understood as the cross-correlation function between the probability density of the photon and the spatial mask analyzer. This cross-correlation provides the probability that the photon is transmitted as a function of the analyzer's transverse position. In other words, the spatial mask \eqref{Mdef} acts as a filter that is used to analyze the spatial structure of the photon field in position \eqref{TProbX} or momentum \eqref{TProbP} domain. It is worth mentioning that since the considered analyzer is an amplitude mask, no phase-sensitive measurements are necessary in the characterization of \eqref{TProb}. This is in contrast with mode analyzers based on spiral \cite{Beijersbergen94,Oemrawsingh04,Oemrawsingh05} or multi-sector \cite{Pors08a,Pors11,Giovannini12} {\it phase} masks that have been utilised in measurements of multi-dimensional orbital-angular-momentum entanglement.

\section{Uncertainty Relation based on periodic analyzers}
\label{sec:UR}
As it is known, the Fourier transform \eqref{OpticalFourier} implies that the photon's transverse field distribution cannot be arbitrarily well localized in both focal planes of the lens. In this paper we explore this complementarity to build an uncertainty relation based on the transmission probabilities of the photon through periodic apertures. To this end, it is convenient to use the Fourier decomposition  of the periodic analyzer \eqref{MFourier} to write the transmission probabilities \eqref{TProb} in the following form:

\begin{subequations}\label{TProbCh}
\begin{equation}\label{TProbChX}
\mathcal{P}(\xi_x)=\sum_{n \in \mathbb{Z}} c_n(\tau_x s_x) \Phi(n \tau_x) \,e^{-i n\tau_x \xi_x},\quad \tau_x=\frac{2\pi}{T_x} ,
\end{equation}
\begin{equation}\label{TProbChP}
\tilde{\mathcal{P}}(\xi_p)=\sum_{n \in \mathbb{Z}} c_n(\tau_p s_p) \tilde{\Phi}(n \tau_p) \,e^{-i n \tau_p \xi_p},\quad \tau_p=\frac{2\pi}{T_p} ,
\end{equation}
\end{subequations}
where $\Phi(\lambda)=\int_{\mathbb{R}}dxP(x) e^{i\lambda x}$ and $\tilde{\Phi}(\lambda)=\int_{\mathbb{R}}dx\tilde{P}(p) e^{i\lambda p}$ stand for the characteristic functions associated with the probability densities $P(x)$ and $\tilde{P}(p)$, respectively. Using the uncertainty relation for characteristic functions reported in Ref. \cite{Rudnicki16}, it is straightforward to show (see Appendix A) that the cross-correlation functions \eqref{TProb} satisfy the  inequality: 
\begin{equation}\label{UR}
\frac{1}{T_x}\int_{-T_x/2}^{T_x/2} d \xi_x  |\mathcal{P}(\xi_x)|^2 +  \frac{1}{T_p}\int_{-T_{p}/2}^{T_{p}/2} d \xi_p  |\tilde{\mathcal{P}}(\xi_p)|^2 \leqslant  
\mathcal{Q},
\end{equation}
with
\begin{equation}\label{URper}
\mathcal{Q} =  \sum_{n\in \mathbb{Z}} \left[C_n^\mathrm{max} +C_n^\mathrm{min} \mathcal{G}\left(n^2 \tau_x\tau_p  \right)\right]\leq \frac{s_x}{T_x}+\frac{s_p}{T_p},
\end{equation}
where 
\begin{equation}
C_n^\mathrm{max/min}=\max/\min\,\{|c_n(\tau_x s_x)|^2,|c_n(\tau_p s_p)|^2\},
\end{equation} 
and the function $\mathcal{G}(\cdot)$ reads \footnote{In \cite{Rudnicki16}, the uncertainty relation for the characteristic functions (Theorem 1 therein) involves the function $\mathcal{B}(\gamma)$, which in the current notation is equal to $1+\mathcal{G}(\gamma)$.}
\begin{equation}\label{BoundFunc}
\mathcal{G}(\gamma)=2\sqrt{2}\frac{\sqrt{2}-\sqrt{1-\cos(\gamma)}}{1+\cos(\gamma)}-1.
\end{equation}
Note that $0\leq \mathcal{G}(\gamma) \leq 1$.  Moreover, the notation in \eqref{UR} involving the moduli of the probability distributions (real, non-negative) is technically speaking unnecessary. Eq.  \ref{UR} is presented in that manner in order to ease the manipulations of apparently complex formulas \eqref{TProbCh}. 

The upper bound $\mathcal{Q}\equiv\mathcal{Q}(T_x,s_x,T_p,s_p)$ is a function of the periodic analyzer parameters used to probe the field distribution in position and momentum domain. Since the functional dependence of the Fourier coefficients \eqref{cn} is only on the ratio $T/s$, the bound function \eqref{URper} turns into a simpler form whenever working with a symmetric arrangement [see Eq. \eqref{URperSym}].
To shortly summarise, the cross-correlation between the periodic analyzer and the transverse field structure of a single photon measured in complementary domains is subject to the uncertainty relation given in \eqref{UR}. This uncertainty relation constitutes the first theoretical result of the paper.

\subsection{Coarse Graining and Faithful Sampling}

Real-world measurements of a CV degree-of-freedom are inevitably subjected to the coarse graining imposed by detection resolution \cite{Rudnicki12a}.
Here the resolution limitation appears due to the necessarily discretised transverse displacements in the experimental characterisation of the cross-correlation functions {\eqref{TProb} -- instead of its (unfeasible) sampling over a continuum of displacements $\xi_{x(p)}$.
Realistically, the experimentalist holds discretised distributions $\mathcal{P}_k \equiv \mathcal{P}(k\Delta\xi_x)$ and $\tilde{\mathcal{P}}_k \equiv \tilde{\mathcal{P}}(k\Delta\xi_p)$ by employing a finite set of displacements with scanning resolution $\Delta\xi_{x}$ and $\Delta\xi_{p}$, respectively. To sample the full periodicity of the analyzer, say in position domain, a number $N_x=T_x/\Delta\xi_{x}$ of transmission probability measurements are required ($k=0, 1, \cdots,N_x-1$). Here we argue that the minimal resolution expected for a ``faithful" sampling is such that $\Delta\xi_{x} \leqslant s_x$: a displacement step greater than the analyzer bin width is unable to  sample the entire field distribution. Interestingly, when using an analyzer geometry with an integer $d =T_x/s_x>1$, the discretised cross-correlation function obtained in the limiting case $\Delta\xi_{x} = s_x$ corresponds to the {\it periodic coarse-graining} \cite{Tasca18a} of the single-photon detection probability distribution. In this case, the set of analyzer displacements corresponds to mutually exclusive outcomes of a $d$-dimensional measurement with $N_x=d$.

\section{Entanglement criteria}
\label{sec:EntCriteria}

In this section we aim to derive experimentally convenient entanglement criteria along the lines discussed in the previous section, {\it i.e.}, based on transmission probabilities through periodic analyzers acting as spatial filters in position and momentum domains. We consider bipartite entanglement present in the spatial degrees-of-freedom of a photon pair. In this scenario, the relevant transmission probability is that of the {\it photon pair}, serving as an indicative of the spatial correlations. 

We name $\varrho_{AB}$ the quantum state associated with the transverse spatial structure of photons $A$ and $B$, and restrict ourselves to the analysis of their spatial structure in a single transverse degree-of-freedom $x$. The generalization to both transverse dimensions is straightforward. Following the definitions of the previous section, we consider the detection probabilities in the conjugate planes of lenses placed in the path of the photons $A$ and $B$. Explicitly, we define the {\it joint} probability distributions:
\begin{subequations}\label{PABvec}
\begin{eqnarray}\label{PABvecpos}
P(\vecx{x}_A,\vecx{x}_B) =& \langle \vecx{x}_A |\otimes \langle \vecx{x}_B | \varrho_{AB} | \vecx{x}_A  \rangle \otimes  |\vecx{x}_B \rangle, \\ \label{PABvecmom}
\tilde{P}(\vecx{p}_A,\vecx{p}_B) =& \langle \vecx{p}_A |\otimes  \langle \vecx{p}_B | \varrho_{AB} | \vecx{p}_A  \rangle \otimes  |\vecx{p}_B \rangle. 
\end{eqnarray}
\end{subequations}
As before, we look at the marginals along the relevant degree of freedom that is probed by the analyzer's periodic aperture:
\begin{subequations}\label{PAB}
\begin{align}\label{PABpos}
P(x_A,x_B) =& \int_{\mathbb{R}}dy_A\int_{\mathbb{R}}dy_B P(\vecx{x}_A,\vecx{x}_B), \\ \label{PABmom}
\tilde{P}(p_A,p_B) =& \int_{\mathbb{R}}dp_{y_A}\int_{\mathbb{R}}dp_{y_B} \tilde{P}(\vecx{p}_A,\vecx{p}_B). 
\end{align}
\end{subequations}
The distributions defined in Eqs. \eqref{PAB} describe spatial correlations of the photons in position and momentum domain, respectively. The probabilities that both photons are transmitted through identical analyzers placed in their propagation path are indicative of the shape and strength of the correlations in question. Our entanglement criteria shall thus be based on such transmission probabilities as quantifiers of the position and momentum correlations shared by the two-photon state $\varrho_{AB}$. The relevant quantities are clearly given by the two-variable counterparts of the cross-correlation functions \eqref{TProb}:
\begin{subequations}\label{JointT}
\begin{equation}\label{JointTpos}
\mathcal{P}(\xi_{x_A},\xi_{x_B})= \int_{\mathbb{R}} \!\!d x_A \int_{\mathbb{R}}  \!\!d x_B \, P(x_A,x_B)  M_{AB}(x_A,x_B),
\end{equation}
\begin{equation}\label{JointTmom}
\tilde{\mathcal{P}}(\xi_{p_A},\xi_{p_B})= \int_{\mathbb{R}} \!\!d p_A \int_{\mathbb{R}} \!\! d p_B \, \tilde{P}(p_A,p_B) M_{AB} (p_A,p_B),
\end{equation}
\end{subequations}
where
\begin{equation}\label{MaskProduct}
M_{AB}(z_A,z_B)=M(z_A - \xi_{z_A};T_z,s_z)M (z_B - \xi_{z_B};T_z,s_z).
\end{equation}

In the two-photon case, the joint distributions $P(x_A,x_B)$ and $\tilde{P}(p_A,p_B)$ are probed by the product of the individual periodic apertures  \eqref{MaskProduct} each photon is subjected to. Obviously, the maximal amount of information on the two-photon correlations is achieved by complete characterization of the cross-correlation functions given in Eqs. \eqref{JointT}. This complete characterization requires the evaluation of the joint transmissions for {\it all} possible values of the transverse displacement of the analyzers used for photons $A$ and $B$. Nevertheless, as we shall now show, the use of periodic apertures allows for entanglement verification with the evaluation of the cross-correlation functions \eqref{JointT} along single directions in phase-space.

We begin by defining $\mathcal{P}_{\pm}(\xi_{x})\equiv\mathcal{P}(\xi_x,\pm \xi_x)$ as the joint transmission probabilities obtained by displacing the apertures for photons $A$ and $B$ along the same ($+$) or opposite ($-$) directions over the full periodicity of the apertures. In other words, $\mathcal{P}_{\pm}$ encode the cross-correlation function \eqref{JointTpos} along the diagonal or anti-diagonal direction of the parameter space given by the individual displacements of the analyzers. Analogously, we define a similar quantity in the momentum domain as $\tilde{\mathcal{P}}_{\pm}(\xi_p)\equiv \tilde{\mathcal{P}}(\xi_p,\pm \xi_p)$. We further introduce:
\begin{subequations}\label{AverageT}
\begin{equation}\label{AverageTpos}
\langle\mathcal{P}_{\pm} \rangle= \frac{1}{T_x} \int_{-T_x/2}^{T_x/2} d\xi_x \, \mathcal{P}(\xi_x, \pm \xi_x),
\end{equation}
\begin{equation}\label{AverageTmom}
\langle \tilde{\mathcal{P}}_{\pm} \rangle= \frac{1}{T_p} \int_{-T_p/2}^{T_p/2}  d\xi_p  \, \tilde{\mathcal{P}}(\xi_p,\pm\xi_p),
\end{equation}
\end{subequations}
as the averages of $\mathcal{P}_{\pm}(\xi_x)$ and  $\tilde{\mathcal{P}}_{\pm}(\xi_p)$, respectively. In Fig. \ref{Fig:ExpSetup}(b) we illustrate the transverse displacement procedure required for the evaluation of the average joint transmission probabilities \eqref{AverageT}. We are now in position to establish our second theoretical result, namely the entanglement criteria. The sum of the average transmission probabilities
\begin{equation}\label{Criteria}
 \langle\mathcal{P}_{\pm} \rangle + \langle \tilde{\mathcal{P}}_{\mp} \rangle \leqslant \mathcal{Q}(T_x,s_x,T_p,s_p)
\end{equation}
is upper bounded by $\mathcal{Q}$ given in Eq. \eqref{URper}, whenever $\varrho_{AB}$ is separable. A violation of \eqref{Criteria} thus implies that $\varrho_{AB}$ is entangled.

{\it Proof:} 
As the starting point we observe that the left hand side of (\ref{Criteria}) is a linear (thus convex) functional of the quantum state, so one can restrict further discussion to separable pure states $\left|\Psi\right\rangle_A \otimes\left|\Psi\right\rangle_B$. In this special case, the joint transmission probabilities factorize: $\mathcal{P}(\xi_{x},\pm \xi_{x})=\mathcal{P}_A(\xi_{x})\mathcal{P}_B(\pm\xi_{x})$. The single-photon probabilities $\mathcal{P}_{A}$ and $\mathcal{P}_{B}$, evaluated for $\left|\Psi\right\rangle_A$ and $\left|\Psi\right\rangle_B$ respectively, are given by Eq. \eqref{TProbX}. The same factorization scheme applies to the momentum domain.

In the second step we use the basic arithmetic--geometric mean inequality, $(a^2+b^2)/2 \geqslant ab$, to separate the $A$ and $B$ parts of the  joint transmission probabilities. More explicitly:
\begin{equation}
\langle\mathcal{P}_{\pm} \rangle \leq  \frac{1}{T_x}\int_{-T_x/2}^{T_x/2} d \xi_x  \frac{|\mathcal{P}_A(\xi_x)|^2 +|\mathcal{P}_B(\pm \xi_x)|^2}{2},
\end{equation}
and similarly for the momentum term. These initial steps are similar to those of Spengler et al \cite{Spengler12}, where entanglement criteria for discrete systems in terms of mutual predictabilities have been derived. 

In fact, the sign combination for the $B$ part and  the labeling of subsystems (pure quantum states) both stop playing a role from now on. The sign ambiguity is removed by the symmetric integrals with respect to $\xi$ while the second assertion is true because in the sum $ \langle\mathcal{P}_{\pm} \rangle + \langle \tilde{\mathcal{P}}_{\mp} \rangle$ we find pairs of probabilities to be bounded by \eqref{UR} independently for each subsystem. This observation concludes the proof.

Due to the symmetry of integrals in \eqref{AverageT}, the same bound as \eqref{Criteria} applies to the even sign combination $ \langle\mathcal{P}_{\pm} \rangle + \langle \tilde{\mathcal{P}}_{\pm} \rangle$. This is not at all surprising, as it is a generic feature of other criteria for continuous variables like those for variances \cite{Duan00,Mancini02} or entropies \cite{Walborn09,Saboia11,Tasca13a}. However, as in the case of the variance and entropic-based inequalities, for the same sign these are true uncertainty relations necessarily obeyed by all bipartite quantum states. At the same time, the corresponding inequalities with different signs need to be satisfied only by separable states. They are also maximally violated by the EPR state, given by normalized versions of  $\psi(x_A,x_B)=\delta(x_A-x_B)$ and $\phi(p_A,p_B)=\delta(p_A+p_B)$.

Even though the grid-geometry provided by the periodic analyzers chosen does not allow for an immediate validation of the first assertion (UR for the same signs), one can easily verify the second one. By a direct calculation one can see that for the EPR state $ \langle\mathcal{P}_{+} \rangle = s_x/T_x$ and $\langle \tilde{\mathcal{P}}_{-} \rangle = s_p/T_p$. The sum of both quantities is equal to the upper bound in Eq. \eqref{URper}.  Thus, for adequate choice of parameters the EPR state violates inequality \eqref{Criteria}. 

\section{Experimental scheme}
\label{sec:experiment}

\begin{figure}[t]
\begin{center}
\includegraphics[width=86mm]{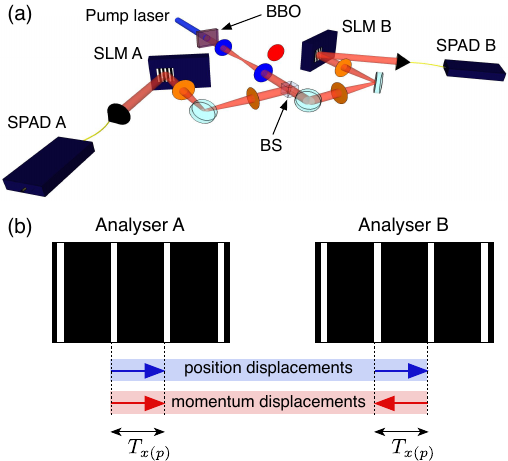}
\caption{(a) Experimental setup. Spatial correlations of entangled photon pairs from SPDC are probed by spatial mask analyzers defined by periodic apertures and displayed on SLMs. The blue (red) lenses are used for measurements in position (momentum) domain. BS: beam splitter; SLM: spatial light modulator; SPAD: single-photon avalanche diode; BBO: Beta Barium Borate non-linear crystal. (b) Illustration of the periodic analyzers for photons $A$ and $B$. The coincidence counts registered while the analyzers are displaced in the same (position) or opposite (momentum) directions are used to test for spatial entanglement.}
\label{Fig:ExpSetup}
\end{center}
\end{figure}

In the experimental part of our investigations we use an SPDC source to generate photon pairs entangled in their transverse spatial variables. As explained in the previous section, we probe their transverse position and momentum correlations using spatial mask analyzers defined by periodic aperture functions. 

Our experimental setup is sketched in Fig. \ref{Fig:ExpSetup}(a). We adjust the SPDC source, pumped by a $325$nm He-Cd laser, to emit frequency-degenerate down-converted photons at $\lambda=650$nm in a collinear configuration. The down-converted photons are split by a $50$:$50$ beam splitter and directed to spatial light modulators (SLM) that are programmed to display the periodic apertures \eqref{Mdef} in the horizontal direction $x$, as illustrated in Fig. \ref{Fig:ExpSetup}(b). 
A switchable lens system placed before the beam splitter is used to produce either the image or the Fourier transform of the SPDC source on to the SLMs. The imaging system is characterised by an optical magnification of $\mathcal{M}=5$, while the optical Fourier transform system has an effective focal length $f_e=333$mm. The optical magnifications both for image plane (IP) and far-field (FF) configurations were chosen in order to maximize the spatial resolution of our measurements while keeping the down-converted beams enclosed  within the SLM panel. Our SLMs (HOLOEYE PLUTO) have an active area of $15.36$mm$\times 8.64$mm with a high-resolution array of $1920 \times 1080$ pixels of width $8\mu$m. 
The photons reflected from the SLMs are lens-coupled to multi-mode optical fibres connected to single-photon avalanche diodes (SPAD). The spatial masks displayed on the SLMs work as aperture functions, reflecting the photons incident upon the white stripes (where $M=1$) and discarding the photons incident upon the black region (where $M=0$). In our experiment, the joint transmission of the photons through the analyzers are measured as coincidence counts  over a sampling time of $1$s. To convert the photo-detection rates to transmission {\it probabilities} we also record the overall joint detection rates. This measurement procedure is realised independently for each configuration (position and momentum) and analyzer geometry tested in our experiment.

\begin{figure}[t]
\begin{center}
\includegraphics[width=86mm]{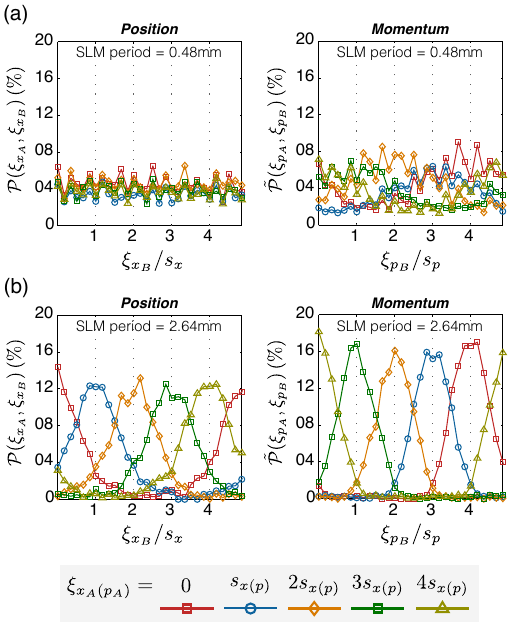}
\caption{Joint transmission probabilities of photon pairs through periodic analyzers measured while scanning analyzer $B$ displacement over a full period of the analyzer. Each set of data points represent a distinct displacement for analyzer $A$, as shown in the legend. The periodicities of the apertures used in the SLM are equivalent to (a) $60$ and (b) $330$ pixels of the SLM. }
\label{Fig:Oscillation}
\end{center}
\end{figure}

To maximise the joint transmission through the periodic analyzers, we perform a calibration procedure to match the relative displacement of the analyzers with the symmetry of the EPR-like correlations displayed by the SPDC photons -- position correlation or momentum anti-correlation \cite{Howell04}. According to the periodic aperture definition \eqref{Mdef}, maximal coincidence counts in position measurements are obtained using a relative displacement of the analyzers given by $\xi_{x_A}-\xi_{x_B}=0$ (mod $T_x$), whereas momentum measurements require $\xi_{p_A}+\xi_{p_B}=T_p-s_p$ (mod $T_p$). The result of this calibration can be visualised in Fig. \ref{Fig:Oscillation}, where we plot the measured transmission probability of the photons as a function of analyzer's $B$ displacement. All displayed measurements were acquired using an analyzer geometry given by $T_x/s_x=T_p/s_p=5$, and each set of data points (plotted in different colours) represent five different displacements of analyzer $A$. Note that in the symmetric case, for which  $T_x/s_x=T_p/s_p=d$, the upper bound \eqref{URper} derived in this paper simplifies to the form
\begin{equation}\label{URperSym}
\mathcal{Q}_\mathrm{sym} =  \sum_{n\in \mathbb{Z}} \left|c_n(2\pi/d) \right|^2\left[1+ \mathcal{G}\left(n^2 \tau_x\tau_p  \right)\right].
\end{equation}
In the measurements presented in Fig. \ref{Fig:Oscillation} (a) we used a periodicity of $0.480$mm ($60$ pixels) to define the apertures displayed on the SLMs. This SLM periodicity led to analyzer periodicities --related to the position and momentum variables at the source-- of $T_x=0.480$mm$/\mathcal{M}=0.096$mm and $T_p=0.480$mm$/\alpha=13.92$mm$^{-1}$. In this case the joint transmission of the photons through the periodic apertures were independent of the analyzer displacement. In other words, these periodicities lead to bin widths $s_x$ and $s_p$ that are too small to capture the spatial correlations of the photons.  On the other hand, the joint transmission measurements plotted in \ref{Fig:Oscillation} (b) present peaks that depend on the displacement of analyzer A.  This type of conditional displacment is somewhat typical in experiments that exploit the EPR-type correlation of the photons. In this latter case, an SLM periodicity $5.5$ times larger was used.

Over the scanning distance of one analyzer period, one coincidence peak is expected for each set of data points, the position of which depends on the displacement of the analyzer $A$ and the symmetry of the correlations displayed at the detection plane (position or momentum). As seen in plots (b) of Fig. \ref{Fig:Oscillation}, the transmission peak for position measurements occurs whenever $\xi_{x_B}=\xi_{x_A}$, whilst the peak transmission for momentum occurs for $\xi_{x_B}=-\xi_{x_A}+4s_p$ (since $T_p=5s_p$ in these measurements). The strength of the spatial correlations is inferred from the visibility of the peaks, whose shape is  approximately a triangular function resembling the auto-correlation function of a slit of width $s_x$ or $s_p$: ideal EPR correlations would generate perfect triangular peaks reaching the maximum transmission probability of $s_{x(p)}/T_{x(p)}=20\%$. Since the five different displacements used for analyzer $A$ are $\xi_{x_A(p_A)}= ks_{x(p)}$ with $k=0,\cdots,4$, we can use the peak values of the transmission probabilities to calculate the average joint transmission probabilities \eqref{AverageT} involved in the entanglement criteria \eqref{Criteria}: 
\begin{eqnarray}
\langle \mathcal{P}_{+} \rangle &=& \frac{1}{d} \sum_{k=0}^{d-1} \mathcal{P}(ks_x, ks_x ), \\
\langle \tilde{\mathcal{P}}_{-} \rangle &=& \frac{1}{d} \sum_{k=0}^{d-1} \tilde{\mathcal{P}}(ks_p, T_p-s_p -ks_p),
\end{eqnarray}
where $d=T_x/s_x=T_p/s_p$ is the number of samples ($5$ for the current measurements).
Indeed, the upper bound \eqref{URperSym} of inequality \eqref{Criteria} calculated for the analyzer parameters used in these measurements [Fig. \ref{Fig:Oscillation}-(b)] limits the sum of the {\it average} transmission probabilities in position and momentum domain to $\mathcal{Q}_\mathrm{sym}=27.18\%$.  Our measurements provide averages of $\langle \mathcal{P}_{+} \rangle=(12.3 \pm0.3 )\%$ and $\langle\tilde{\mathcal{P}}_{-}\rangle=(16.6 \pm0.3 )\%$ for position and momentum, respectively, leading to a violation of the criteria \eqref{Criteria},  
\begin{equation}\label{ViolationExample}
\langle \mathcal{P}_{+} \rangle+\langle\tilde{\mathcal{P}}_{-}\rangle=(28.9 \pm0.4 )\% \nleqslant 27.18\%,
\end{equation}
by over $4$ standard deviations. The presented errors are calculated assuming Poissonian statistics for the measured coincidence counts.

\begin{figure}[t]
\begin{center}
\includegraphics[width=86mm]{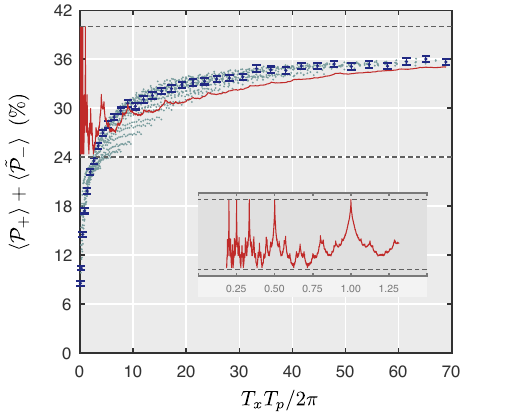}
\caption{Experimental data (blue dots) obtained for the sum of the average joint transmission of photon pairs through periodic analyzers positioned in the image plane (position) and far-field (momentum) of the source. The used analyzer geometry is a symmetric arrangement with $d=5$. The red curve indicates the bound function \eqref{URperSym}. Spatial entanglement is detected whenever $\langle \mathcal{P}_{+} \rangle+\langle\tilde{\mathcal{P}}_{-}\rangle>\mathcal{Q}_\mathrm{sym}$. The inset shows the oscillating structures of $\mathcal{Q}_\mathrm{sym}$ for small values of $T_xT_p/2\pi$.}
\label{Fig:BoundAnalysis}
\end{center}
\end{figure}

\section{Experimental Results}
\label{sec:results}

Prompted by the successful experimental verification of spatial entanglement exemplified in \eqref{ViolationExample}, we now examine the effectiveness of the derived entanglement criteria \eqref{Criteria} for a wider set of periodic analyzer geometries. In the symmetric case, as studied in the last section, the function \eqref{URperSym} upper-bounding the sum of the average transmission probabilities depends solely on the product of the analyzer periodicities ($T_xT_p$) and on the ratio $d=T_x/s_x=T_p/s_p$. The red solid line in Fig. \ref{Fig:BoundAnalysis} displays $\mathcal{Q}_\mathrm{sym}$ for $d=5$ as a function of $T_xT_p/2\pi$. Rather than a simple shape, $\mathcal{Q}_\mathrm{sym}$ presents an intricate functional dependence on $T_xT_p$, including an oscillatory structure for $T_xT_p/2\pi \leqslant1$ (see inset). Interestingly, $\mathcal{Q}_\mathrm{sym}$ assumes its maximal possible value of $2/d$ [see Eq. \eqref{URper}] whenever $T_xT_p=2\pi/m$ for all $m\in \mathbb{N}$, shown as the upper dashed line for $d=5$ in Fig. \ref{Fig:BoundAnalysis}. This is in full agreement with the fact that compatible observables do exist in periodic coarse-graining structures with these particular combinations of periodicities in position and momentum domains \cite{Tasca18a,Busch86,Reiter89}. Obviously, no useful entanglement tests exist in this case. On the other hand, the lower dashed line indicates the value $\langle \mathcal{P}_+ \rangle+\langle \tilde{\mathcal{P}}_-\rangle= 1/d+1/d^2$ that represents the average joint transmission achieved with perfect correlation in one domain ($1/d$) and the absence of correlation in the conjugate domain ($1/d^2$). This construction of maximal correlations in only one basis is typical from entanglement criteria for quantum systems of finite dimension \cite{Spengler12}. As seen from the inset, the lowest values attained for $\mathcal{Q}_\mathrm{sym}$ do not exactly reach this minimum in general (we find that $\mathcal{Q}_\mathrm{sym}$ does saturate to this minimum for $d=2$), but lie slightly above. 

Our experimental data are plotted as blue dots in Fig. \ref{Fig:BoundAnalysis}. Using the SPDC setup described in the last section, we recorded a series of transmission probability measurements while varying the analyzer periodicities displayed on the SLMs  from $0.24$mm up to $8.64$mm. The dark-blue data points with error bars correspond to position and momentum measurements taken with matching SLM periodicities (in IP and FF configurations), whilst the light-blue dots represent the sum of the average transmission probabilities for all other combination of analyzer periodicities. As expected, both $\langle \mathcal{P}_+\rangle$ and $\langle \tilde{\mathcal{P}}_-\rangle$ increase as a function of the analyzer periodicity, leading to violations of the entanglement criteria for sufficiently large periods. Also, we obtain $\langle \mathcal{P}_+\rangle \approx \langle \tilde{\mathcal{P}}_-\rangle \approx 1/d^2$ when using  the smallest SLM periodicity of $0.24$mm. This is an evidence that the correlation width of the photons \cite{Schneeloch16} is greater than (or of the order of) the associated bin width, which at the SLM correspond to $48\mu$m (with $d=5$). 
Evidently, the EPR-like spatial correlations of the photons are not sharp enough to allow for entanglement verification with the product $T_xT_p$ lying in the range shown in the inset of Fig. \ref{Fig:BoundAnalysis}.

\begin{figure*}[t]
\begin{center}
\includegraphics[width=160mm]{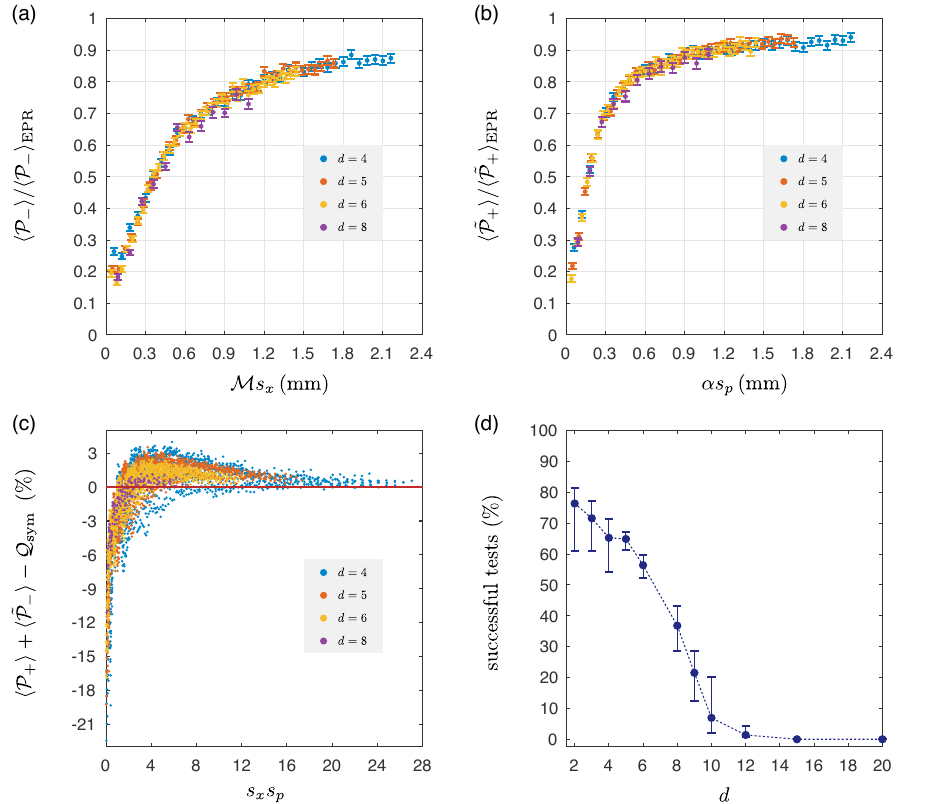}
\caption{Normalized average transmission probabilities for analyzer geometries $d=4,5,6,8$ for (a) position (near-field) measurements and (b) momentum (far-field) measurements as a function of the physical bin width of the periodic analyzers.  (c) Entanglement test for average transmission probabilities obtained from data points shown in (a) and (b). One can see that entanglement is best observed near $s_xs_p \approx 4$. (d) Overall success probability for entanglement detection as a function of the analyzer parameter $d$, using data corresponding to all bin widths (see text for more details).}
\label{Fig:AnalysisBinWidth}
\end{center}
\end{figure*}

The optimal analyzer geometry leading to successful entanglement verification is highly dependent on the spatial  correlations; both the periodicity and the bin width can be adjusted to optimise entanglement detection.  For the EPR-like spatial correlations generated in our experiment, the most relevant parameter to maximise the joint transmission is the bin width of the periodic analyzer. To show this, we compare $\langle \mathcal{P}_+\rangle$ and $\langle \tilde{\mathcal{P}}_-\rangle$ measured for four different analyzer geometries ($d=4,5,6$ and $8$) as a function of the bin width. Since the maximal value allowed also depends on $d$, we normalise the measured quantities with those associated with ideal EPR correlations: $\langle \mathcal{P}_+\rangle_{\mathrm{EPR}}=\langle \tilde{\mathcal{P}}_-\rangle_{\mathrm{EPR}}=1/d$. As seen in plots (a) and (b) of Fig. \ref{Fig:AnalysisBinWidth}, the same trend is observed for the average joint transmissions as a function of the bin width, regardless of the parameter $d$. It is also possible to recognise that the far-field correlations in our setup are stronger than those in the image plane of the source (note that the plots abscissae are the physical bin widths of the apertures displayed on the SLMs). Since the largest periodicity applied to the SLMs ($8.64$mm) was the same for all analyzer geometries, the largest used bin width is smaller for larger $d$.

Each data point in plot (c) of Fig. \ref{Fig:AnalysisBinWidth} represents an entanglement test calculated from the average transmission probabilities displayed in plots (a) and (b). For these tests, we mixed only position and momentum data with the same $d$, thus corresponding to a symmetric arrangement of analyzers. Entanglement is successfully detected according to our criteria for every data point lying above the red horizontal line representing $\langle \mathcal{P}_+\rangle+\langle \tilde{\mathcal{P}}_-\rangle - \mathcal{Q}_\mathrm{sym}=0$. An optimal range of analyzer parameters leading to the greatest violations of inequality \eqref{Criteria} is centred around $s_xs_p \approx 4$. This product value corresponds, for example, to position and momentum data taken with the same SLM bin width of $\mathcal{M}s_x=\alpha s_p\approx0.83$mm. These bin widths are wide enough to allow a large joint transmission of the correlated photons, but not so large as to unnecessarily increase $\mathcal{Q}_\mathrm{sym}$ via its dependence on the product $T_xT_p=d^2s_xs_p$. As a consequence, the success rate of entanglement detection as well as the largest achieved violation decrease with $d$. The success rate of entanglement tests in our experiment is shown in plot (d) of Fig. \ref{Fig:AnalysisBinWidth}. These success rates were calculated from $1296$ entanglement tests for each $d \leqslant 6$ and $144$ tests for each $d > 6$. The largest SLM periodicity used in these tests was the same for all $d$.
The lower part of the error bars represent violations by less than one standard deviation [the error as in Eq. \eqref{ViolationExample}], whilst the upper part accounts for unsuccessful tests by less than one standard deviation. In total, we were able to detect entanglement in $4432$ out of $7344$ tests, thus achieving an overall success rate of about $60\%$.

\section{Discussion}
\label{sec:conclusion}

We have demonstrated that the entanglement present in a bipartite continuous variable system can be detected using a special kind of coarse-graining measurement using periodic sampling projectors. We illustrate the usefulness of the method with an experiment using twin photons from parametric down-conversion entangled in transverse spatial variables. The results have shown that the method works very well and can be easily extended to other CV systems. In comparison with the traditional method of binning the measurement variable, our method presents the advantage of working with higher signal to noise ratio. This advantage comes from the periodicity of the used analyzers that allows the light to be transmitted through more than one slit, while in the standard binning only one slit is allowed. Moreover, in order to achieve enough resolution, traditional binning usually requires very thin slits. In contrast with this requirement, our method was shown to prove entanglement even using slit widths that would be considered very large for the standard binning procedure.

The reported criteria may be a convenient and robust test of entanglement present in non-trivial CV states that itself present periodic phase space structures \cite{Neves05,Carvalho12,Barros17}.
It is worth noting that our results are applicable to any quantum continuous variable, and can be used even if a complete CV measurement has been made, as it is possible to post-process the measured distributions into the necessary periodic binning structure. 

\begin{acknowledgements}
D.S.T would like to thank D. Giovannini, M. Krenn, G. H. Aguilar, O. J. Farias and F. de Melo for stimulating discussions. This research was funded by the Brazilian funding agencies FAPERJ, CAPES, CNPq and the National Institute for Science and Technology - Quantum Information. {\L}.R. acknowledges financial support by grant number 2014/13/D/ST2/01886 of the National Science Center, Poland.  M.J.P. and R.S.A acknowledge financial support from the UK EPSRC under a Programme Grant, COAM award number EP/I012451/1.
\end{acknowledgements}

\appendix

\section{Concise proof of the uncertainty relation for periodic analyzers} \label{appendix:UR_proof}
The formula \eqref{TProbChX} based on the Fourier expansion of the spatial mask function trivially leads to ($\Delta n=n'-n$):
\begin{equation}\label{TPmod2}
|\mathcal{P}(\xi_x)|^2\!=\!\!\!\!\sum_{n,n' \in \mathbb{Z}}\!\!\! c_n(\tau_x s_x)   c_{n'}^*(\tau_x s_x)\Phi(n \tau_x)\Phi^*(n' \tau_x) e^{i \Delta n\tau_x \xi_x}.
\end{equation}
The integral with respect to $d\xi_x$ present in \eqref{UR} shall only involve the last factor, namely $ e^{i \Delta n\tau_x \xi_x}$. Since we integrate over the whole period, the integration provides $T_x \delta_{n'n}$. With this result we can perform the sum with respect to $n'$ and obtain
\begin{equation}\label{IntTPmod2}
\frac{1}{T_x}\int_{-T_x/2}^{T_x/2} d \xi_x  |\mathcal{P}(\xi_x)|^2 =\sum_{n \in \mathbb{Z}}|c_n(\tau_x s_x)|^2 |\Phi(n \tau_x)|^2.
\end{equation}
The same line of reasoning applies to the second term associated with the momentum variable. As a result, we find that the left hand side of \eqref{UR} is equal to
\begin{equation}
\sum_{n \in \mathbb{Z}}\left[|c_n(\tau_x s_x)|^2 |\Phi(n \tau_x)|^2+|c_n(\tau_p s_p)|^2 |\tilde\Phi(n \tau_p)|^2\right].
\end{equation}
In the last step, we maximize the above quantity with respect to both characteristic functions, independently for each value of $n$. The optimization procedure is subject to the constraint
\begin{equation}
|\Phi(n \tau_x)|^{2}+|\tilde\Phi(n \tau_p)|^{2}\leq 1+ \mathcal{G}(n^2\tau_x\tau_p),
\end{equation}
which is a particular instance of the uncertainty relation for the characteristic functions derived in \cite{Rudnicki16}. Also, from the sole definition of the characteristic function, we know that $|\Phi(n \tau_x)|^{2}\leq 1$ and $|\tilde\Phi(n \tau_p)|^{2}\leq 1$. In mathematical terms, we are thus maximizing the function $a X+b Y$ with respect to $(X,Y)$. All $a,b,X,Y$ are real and non-negative, and we have the constraints: $X\leq 1$, $Y\leq 1$ and $X+Y\leq 1+\mathcal{G}$ with $0 \leq \mathcal{G}\leq 1$. Since the above inequalities define a convex set, the maximum is obtained at the edge points. If $a\leq b$ then the maximizing edge is $Y=1$ and $X=\mathcal{G}$, while when  $a\geq b$ we shall pick up $X=1$ and $Y=\mathcal{G}$. As a result (considering both cases together) we obtain the first desired inequality (exact formula for $\mathcal{Q}$).

The second inequality part pops up after one bounds $ \mathcal{G}(n^2\tau_x\tau_p)$ by $1$. Then
\begin{equation}\label{H1}
\mathcal{Q}\leq \sum_{n \in \mathbb{Z}}\left[|c_n(\tau_x s_x)|^2 +|c_n(\tau_p s_p)|^2 \right].
\end{equation}
Since within a single period, the transmittance $|M(x)|^2$ is equal to $1$ for $0\leq x   {\rm \,(mod \,}  T {\rm)} \leq s$ and vanishes elsewhere, we obtain the relations:
\begin{equation}
\frac{s}{T}=\frac{1}{T}\int_{0}^{T} d x |M(x)|^2 = \sum_{n \in \mathbb{Z}}|c_n(\tau  s)|^2,
\end{equation}
which turns (\ref{H1}) into the inequality part of (\ref{URper}).




\end{document}